\documentclass[a4papper]{article}
\usepackage[dvips]{color}
\usepackage{graphicx}
\usepackage{graphics}

\usepackage[cp850]{inputenc}
\usepackage[spanish,english]{babel}
\usepackage{amsfonts,amsbsy,amsmath}
\usepackage{epsfig}
\usepackage{multicol}
\usepackage{colortbl}
\usepackage{pstricks,pst-grad}

\input epsf

\begin{document}

\begin{figure}
\begin{center}
\resizebox{1\textwidth}{!}{ \includegraphics{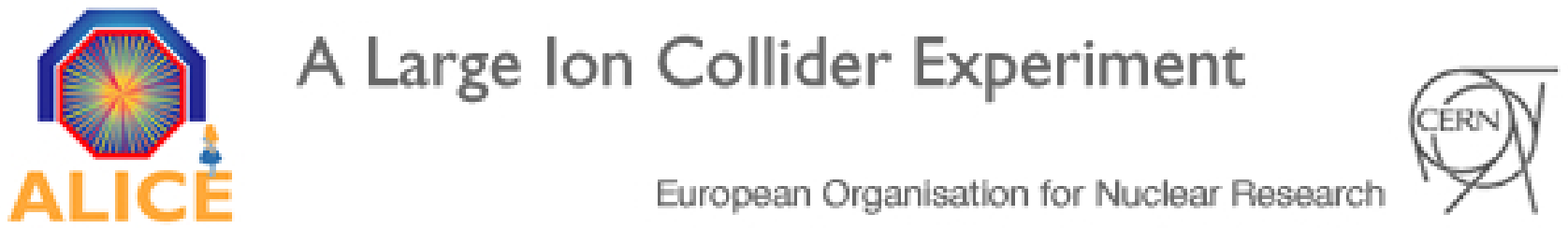} % Please
% \includegraphics[width=10cm,height=11cm]{TvsRESDMCjetjet100300109.eps}
  %add fig...  }
  }%\vspace{0.2cm}
  %\caption{Shape variables at hadron colliders are defined over
%particles within a central region, $C$, and by the addition of a
%recoil term also defined within $C$, are sensitive to the forward
%region $\overline{C}$.} \label{central}
\end{center}
\end{figure}

\title{Event Shape Analysis in ALICE}
\author{Antonio Ortiz Velasquez $^{(1)}$, Guy Pai\'c $^{(1)}$
                                   \\
$^{(1)}$ Instituto    de    Ciencias Nucleares, Universidad
Nacional  Aut\'onoma  de  M\'exico, \\
  Apartado Postal   70-543,   M\'exico    D.F.   04510,
  M\'exico`.}
\maketitle{}

\vspace{2.0cm}

\flushright{Internal Note/Physics \\
ALICE reference number \\
ALICE-INT-2009-015 version 1.0}

\vspace{3.0cm}

\abstract{ The jets are the final state manifestation of the hard
parton scattering. Since at LHC energies the production of hard
processes in proton-proton collisions will be copious and varied,
it is important to develop methods to identify them through the
study of their final states. In the present work we describe a
method based on the use of some shape variables to discriminate
events according their topologies. A very attractive feature of
this analysis is the possibility of using the tracking information
of the TPC+ITS in order to identify specific events like jets.
Through the correlation between the quantities: thrust and recoil,
calculated in minimum bias simulations of proton-proton collisions
at 10 TeV, we show the sensitivity of the method to select
specific topologies and high multiplicity. The presented results
were obtained both at level generator and after reconstruction. It
remains that with any kind of jet reconstruction algorithm one
will confronted in general with overlapping jets. The present
method determines areas where one does encounter special
topologies of jets in an event. The aim is not to supplant the
usual jet reconstruction algorithms, but rather to allow an easy
selection of events allowing then the application of
algorithms. }

\newpage
%\vspace{0.3cm}
\tableofcontents

%\vspace{2cm}

\newpage
\section{Introduction}
In QCD the jets are defined as cascades of consecutive emissions
of partons initiated by partons from any initial hard process. The
partons produce the observed hadrons due to confinement.  Di-jets
were discovered in 1975 in $e^{+}e^{-}$ collisions \cite{hanson},
the observation of three coplanar jets has provided the first
experimental evidence for the existence of gluon
\cite{Barber:1979yr, pluto, tasso, jade}.

Event shapes measure geometrical properties of the energy flow in
QCD states. Especially in the case of $e^{+}e^{-}$ collisions and
Deep Inelastic Scattering, they were among the most studied QCD
observables, both theoretically and experimentally. Sphericity was
used at SLAC to show the evidence for the existence of jets in the
annihilation process:  $e^{+}e^{-}$ at energies up to 7.4 GeV in
the c. m. \cite{hanson}. In 1979 the collaboration MARK-J used the
variable oblateness for describing processes where three prolong
jets were produced at energies up to 31.5 GeV in the c. m.
\cite{Barber:1979yr}.

Of course the shape variables are calculated in terms of the final
partons and they are measured in terms of the final observed
hadrons, so it is possible to use them in order to do studies of
corrections by hadronization effects \cite{2,3}. Also a vast
number of strong coupling constant measurements was reported
\cite{1}. In the case of the MC generators the validation of some
results were also published \cite{3-1,3-2}.

Experimentally, jets in proton-proton collisions are defined as an
excess of transverse energy over the background of the underlying
events\footnote{Underlying events are formed from the beam-beam
remnants, initial-state radiation and possible from soft and
semi-hard multiple parton interactions.} with a typical cone
radius $R_{C}=1$ in $\eta - \phi$. In ALICE we do not yet have an
extensive calorimeter. For this reason, in order to define and
reconstruct jets we are using tracking measurements. The jets
identification is based on different algorithms; for example the
cone ones. In this work we propose a method which uses the event
shapes to identify interesting topologies of events like jets,
which of course does not prevent the later use of jet algorithms
on the selected events.

\begin{figure}
\begin{center}
\resizebox{1.0\textwidth}{!}{ \includegraphics{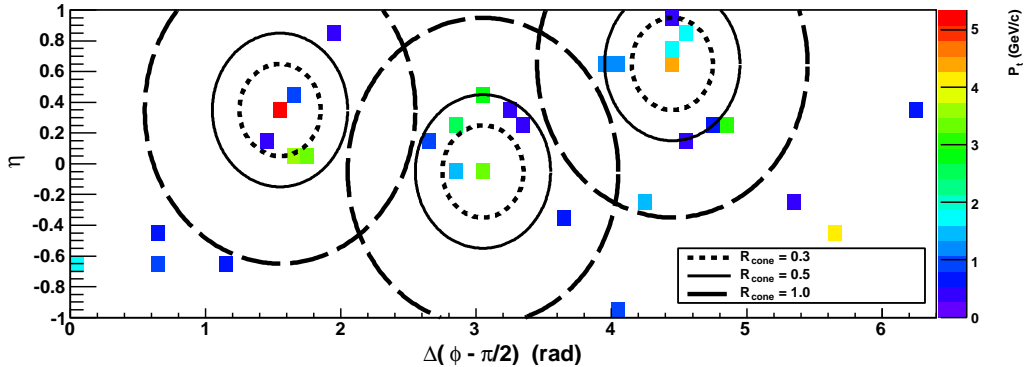} % Please
% \includegraphics[width=10cm,height=11cm]{TvsRESDMCjetjet100300109.eps}
  %add fig...  }
  }\vspace{0.2cm}
  \caption{Plot: $\eta$ vs. $\Delta (\phi-\pi/2)$ vs. $p_{t}$ for a typical event. The leading particle
  is placed at $\pi/2$ rad. Different radius are shown.} \label{overlapping}
\end{center}
\end{figure}

The main problem we will face at LHC is the big number of jets
overlapping  one another. This makes all analysis subject to
numerous cuts  which do increase the systematic uncertainties of
the results. A typical events represented in the eta-phi plane is
shown in Fig. \ref{overlapping}. To guide the eye the contours of
jets with $R_{cone}$=0.3, 0.5 and 1 are shown for the ALICE
acceptance. However in this note we would like to emphasize the
fact that there is a possibility, using the event shapes, to
identify painlessly events where the topology is much simpler than
shown in Fig. \ref{overlapping}. In the present note we first
describe the technicalities of the method of event shape analysis
giving some overview of salient topologies like events where the
jets are only partially detected in the acceptance, events with 2
jets and events with 3 jets. Then, the main features of each class
are studied and demonstrated, like total pt spectra of jets, the
sensitivity on the parameters variation within the method, the
yield of specific topologies in function of multiplicity. Finally
we introduce also a bulk analysis of all events without selection.
We used this to present a comparison between Pythia and PHOJET
events for different parameter values.

\section{Definitions of the shape variables}
By construction the following quantities are Infra Red and
Collinear (IR$\&$C) safe. The shape variables at hadron colliders
are defined over particles within the acceptance of the detector.
The thrust ($T$) is defined as in the $e^{+}e^{-}$ case, but using
only transverse variables \cite{gavin}:
\begin{equation}
T\equiv \underbrace{max}_{\overrightarrow{n}_{t}}
\frac{\sum_{i}|\overrightarrow{p}_{t,i}\cdot\overrightarrow{n}_{t}|}{\sum_{i}|\overrightarrow{p}_{t,i}|}
\end{equation}
In the literature it is more common to find the following definition:
\begin{equation}
\tau\equiv 1-T
\end{equation}
(related to the sphericity  of the event).

The range of $\tau$ is between 0 (for events with narrow
back-to-back jets) and $1/2$ (for events with a uniform
distribution of momentum).

\begin{figure}
\begin{center}
\resizebox{0.6\textwidth}{!}{ \includegraphics{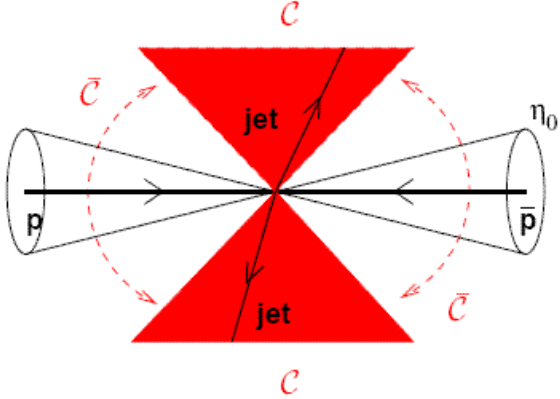} % Please
% \includegraphics[width=10cm,height=11cm]{TvsRESDMCjetjet100300109.eps}
  %add fig...  }
  }\vspace{0.2cm}
  \caption{Shape variables at hadron colliders are defined over
particles within a central region, $C$, and by the addition of a
recoil term also defined within $C$, are sensitive to the forward
region $\overline{C}$.} \label{central}
\end{center}
\end{figure}

The recoil term  $R$ is simply the vector sum of the transverse
momentum:  (Fig. \ref{central})
\begin{equation}
R\equiv \frac{1}{\sum_{i}|\overrightarrow{p}_{t,
i}|}|\sum_{i}\overrightarrow{p}_{t i}|
\end{equation}
This quantity measures the balance of momenta of the event. For
example for a di-jet event, with only one jet inside the
acceptance of the detector (monojet in the further text): $R$ tends to 1, because there are no
vectorial cancellations in the numerator which appear in the
definition of $R$. Otherwise, in the case of the perfect 2
back-to-back jet completely inside the acceptance: $R$ tends to 0.

\section{Description of the event shape analysis (ESA)}
We have computed the shape variables at level generator and
reconstruction. The participants in the computations are primary
charged particles \footnote{Primary particles are defined as
particles produced in the collision, including products of strong
and electromagnetic decays, but excluding feed-down products from
strange weak decays and particles produced in secondary
interactions. In the simulation these are the final state
particles created by the event generator, which are then
propagated (and decayed) by the subsequent detector simulation.}
in the generation case, and  tracks associated with primary
particles in the reconstruction one. In both cases events are
selected using the MB2\footnote{This trigger uses the logical AND
between the V0-OR signal and the Pixel-Fast-OR. And it is an
option if one needs to assign each trigger to a specific bunch
crossing \cite{PPRVolumeII}.} trigger criteria and we restrict the
analysis for events with primary vertex in $z$ direction:
$|v_{z}|\leq10$ cm.

\begin{figure}
\begin{center}
\resizebox{0.8\textwidth}{!}{ \includegraphics{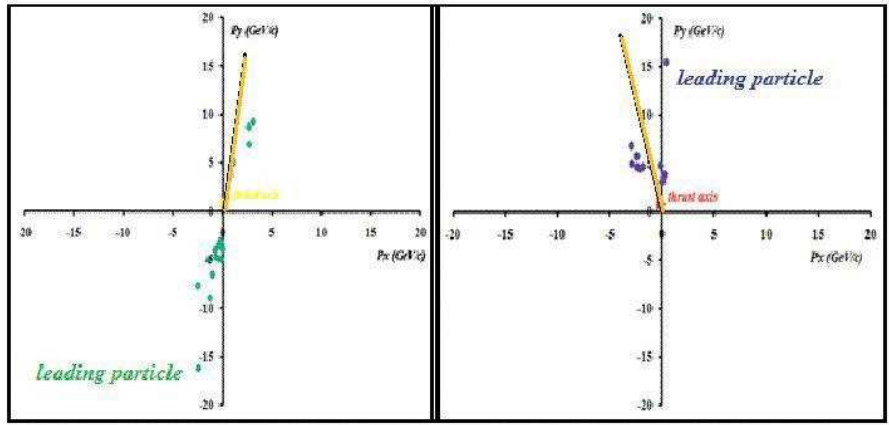} % Please
  }\vspace{0.5cm}
  \caption{Illustration of the distribution of the momenta in the transverse plane for two classes of events. (left) Dijet event, (right) monojet event.
  In both cases the thrust axis is very close to the direction of the leading particle.} \label{ilustraciondijet}
\end{center}
\end{figure}

The requirements to perform the computation are:
\begin{enumerate}
\item \textbf{Generation level:}
\begin{itemize}
\item[Event level:] the first step is for selecting hard events:
$p_{t}^{leading}\geq3$ GeV/c and $|\eta^{leading}|\leq0.5$. This
requirement guarantees that the leading jets are contained within
the ALICE acceptance with a $\Delta
R_{cone}=\sqrt{\Delta\eta^{2}+\Delta\phi^{2}}\leq0.7$

 \item[Particle level:] for primary charged
particles in the acceptance: $|\eta|\leq1$ and $p_{t}\geq1.5$
GeV/c, the shape variables are computed. The cut in $p_{t}$
eliminates the mostly soft underlying component.

\end{itemize}
\item \textbf{Reconstruction level:} In the present analysis we
used tracks reconstructed by the TPC and ITS. The event cuts
described above are applied.

\begin{itemize}
\item[Track level:] tracks associated to primary particles in the
acceptance: $|\eta|\leq1$ and with $p_{t}\geq1.5$ GeV/c.  To
select this class of tracks we applied the following cuts.
\begin{enumerate}
\item TPC refit. \item At least 50 clusters in TPC. \item
Covariance matrix cuts. \item Reject kink daughters.\item Maximum
DCA (in $xy$ and $z$) to vertex $3$ cm.
\end{enumerate}

\end{itemize}
For more details see \cite{janfieteeta}.
\end{enumerate}
In the Fig. \ref{ilustraciondijet} we show a picture of two
classes of candidate events: di-jet and mono-jet (to separate
these events we used their values of shape variables according
with the discussion in the next paragraphs). For a given event we
have taken the projection of the  particles momentum in the
transverse plane. It is well visible that the thrust axis is very
close to the direction of the transverse momenta of the leading
particle (particle with the highest $p_{t}$ in the event).

The method starts by plotting a two dimensional distribution
(``thrust map''), with $\tau$ ($1-T$) in the horizontal axis and
$R$ in the vertical axis. This plot allows to identify different
classes of events according with their location in the thrust map.
\begin{enumerate}
\item \textbf{Region A.}  Suppose a di-jet event which occurs
completely inside the ALICE acceptance ($|\eta|\leq1$). In this
case, we have in the transverse plane; the thrust axis
($\overrightarrow{n}_{t}$) almost collinear to the direction of
the leading particle. So, the ratio of definition (1) tends to 1,
and the recoil term tends to 0. So, the region A is characterized
by events with small values of $1-T$ and $R$, corresponding to
di-jet events.

\item \textbf{Region B.} Events with only one jet in the
acceptance of the detector will have small values of $\tau$ and
due to the absence of vectorial cancellations in the numerator of
the recoil term, these classes of events will have the biggest
values $R$. So,  the region B is populated by monojet events.

\item \textbf{Region C.} The most isotropic events (with high
$1-T$ and small $R$) of the sample have to compensate the
transverse momentum.
%particles (or tracks) with certain cutoff in $p_{t}$, it ispossible to discriminate events with a peculiar three jettopology.
This zone of the thrust map is characterized by the presence of three jet events that we like to call, as at LEP,
 ``mercedes'' events.

\end{enumerate}

The intermediate region between A and B is populated by the
combination of mono-jets and ``incomplete di-jets''; di-jets are
incomplete due to their high $R$, many particles are outside of
the acceptance.

The intermediate region between B and C is populated by events
were 3 or more jets are emitted but at small angles with respect
to the direction of the leading jet.

In the present note we will not deal with these cases which will
be the subject of a special study at later time.

\section{Event shape analysis in minimum bias simulations.}

\begin{figure}
\begin{center}
\resizebox{0.7\textwidth}{!}{ \includegraphics{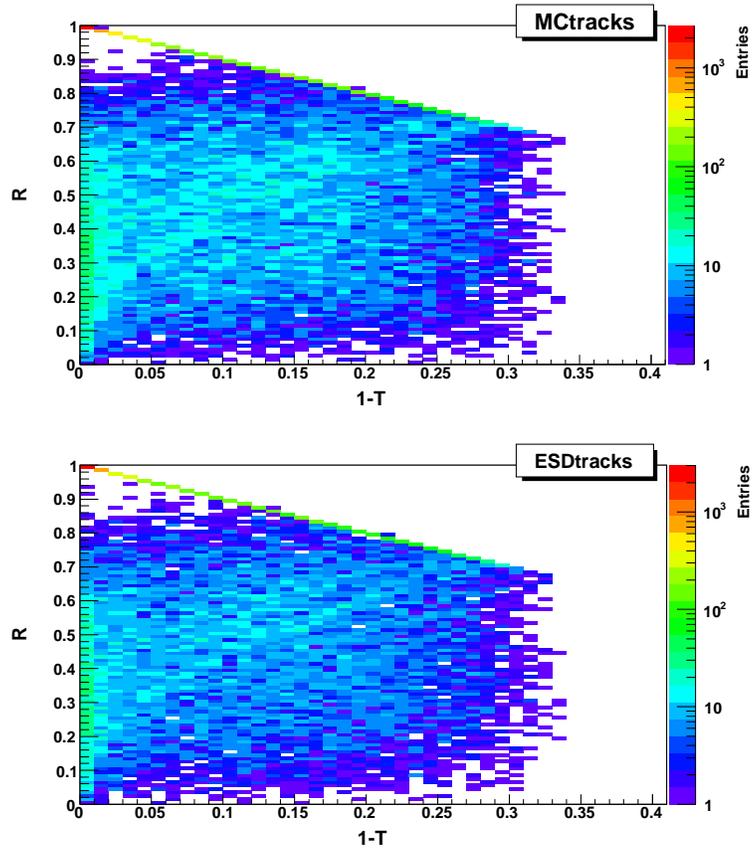} % Please
% \includegraphics[width=10cm,height=11cm]{TvsRESDMCjetjet100300109.eps}
  %add fig...  }
  }%\vspace{0.5cm}
  \caption{Thrust map: a) generator level (upper), in this case we use primary charged MC particles; and b)  reconstruction level
(bottom), in this case we used tracks associated to primaries from
TPC+ITS information.} \label{thrustmapjetjet}
\end{center}
\end{figure}

The present analysis uses the standard simulated samples staged on
alicecaf\footnote{The CERN Analysis Facilities for ALICE
(alicecaf) is a cluster at CERN running PROOF (Parallel ROOT
Facility) which allows interactive parallel analysis on a local
cluster.}. The production corresponds to PDC09, minimum bias
simulations of proton-proton collisions at 10 TeV in the center of
mass, the events were generated with Pythia. The simulation of the
detector included a magnetic field of 0.5 $T$.  The results were
obtained through the analysis of $1,200,000$ events. In the figure
\ref{thrustmapjetjet} the thrust map is shown for generated
(upper) and reconstructed (bottom) data. In both histograms two
interesting regions are exhibited (A and B regions described
before). Note, that any exceptional event with an azimuthally
uniform distribution would be immediately ``detected'' in the
unpopulated area: $1-T\geq0.35$.

 The line which appears in the high R part of the map is due to the
definition of the variables. For example, suppose that we have one
event with a certain thrust axis $\overrightarrow{n}_{t0}$, and in
the event there are $N$ particles. If the vectorial transverse
momentum of the particle $i$ is $\overrightarrow{p}_{t,i}$, then
from the definitions of R and T:
\begin{equation}
1-T+R=1-\frac{\sum_{i}|\overrightarrow{p}_{t,i}\cdot\overrightarrow{n}_{t0}|}{\sum_{i}|\overrightarrow{p}_{t,i}|}+\frac{|\sum_{i}\overrightarrow{p}_{t
i}|}{\sum_{i}|\overrightarrow{p}_{t, i}|}
\end{equation}
Now, we use the fact that:
\begin{equation}
|\sum_{i}\overrightarrow{p}_{t
i}|\leq\sum_{i}|\overrightarrow{p}_{t, i}|
\end{equation}
Then:

\begin{equation}
1-T+R \leq 1-T+1
\end{equation}
The maximum value which $T$ can reach is 1, so we find the
following restriction:
\begin{equation}
1-T+R \leq 1
\end{equation}

\begin{figure}
\begin{center}
\resizebox{0.5\textwidth}{!}{ \includegraphics{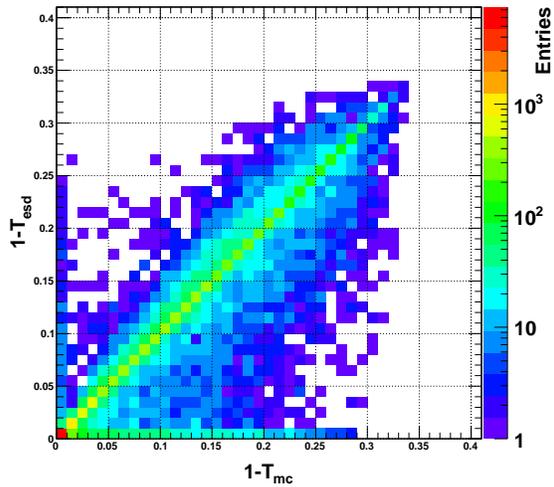} % Please
  }\vspace{0.5cm}
  \caption{Correlation between the quantity $1-T$ computed from MC information and from TPC+ITS information.} \label{thrustjetjetMCESD}
  \end{center}
\end{figure}

In the Fig. \ref{thrustjetjetMCESD} there is a plot which shows
the correlation of the $1-T$ values computed from the generator
information and  from the reconstruction. Note that there is a
small leakage of events with $1-T_{mc}\geq0.03$ to the
reconstructed zone associated to events with 2 back-to-back jet
structure. This can be understood in terms of the reconstruction
effects. For example a generated event with three primary charged
particles ($p_{t}\geq1.5$ GeV/c) distributed isotropically in the
azimuth could be reconstructed as a event with a structure of two
jets.

\begin{figure}
\begin{center}
\resizebox{0.5\textwidth}{!}{ \includegraphics{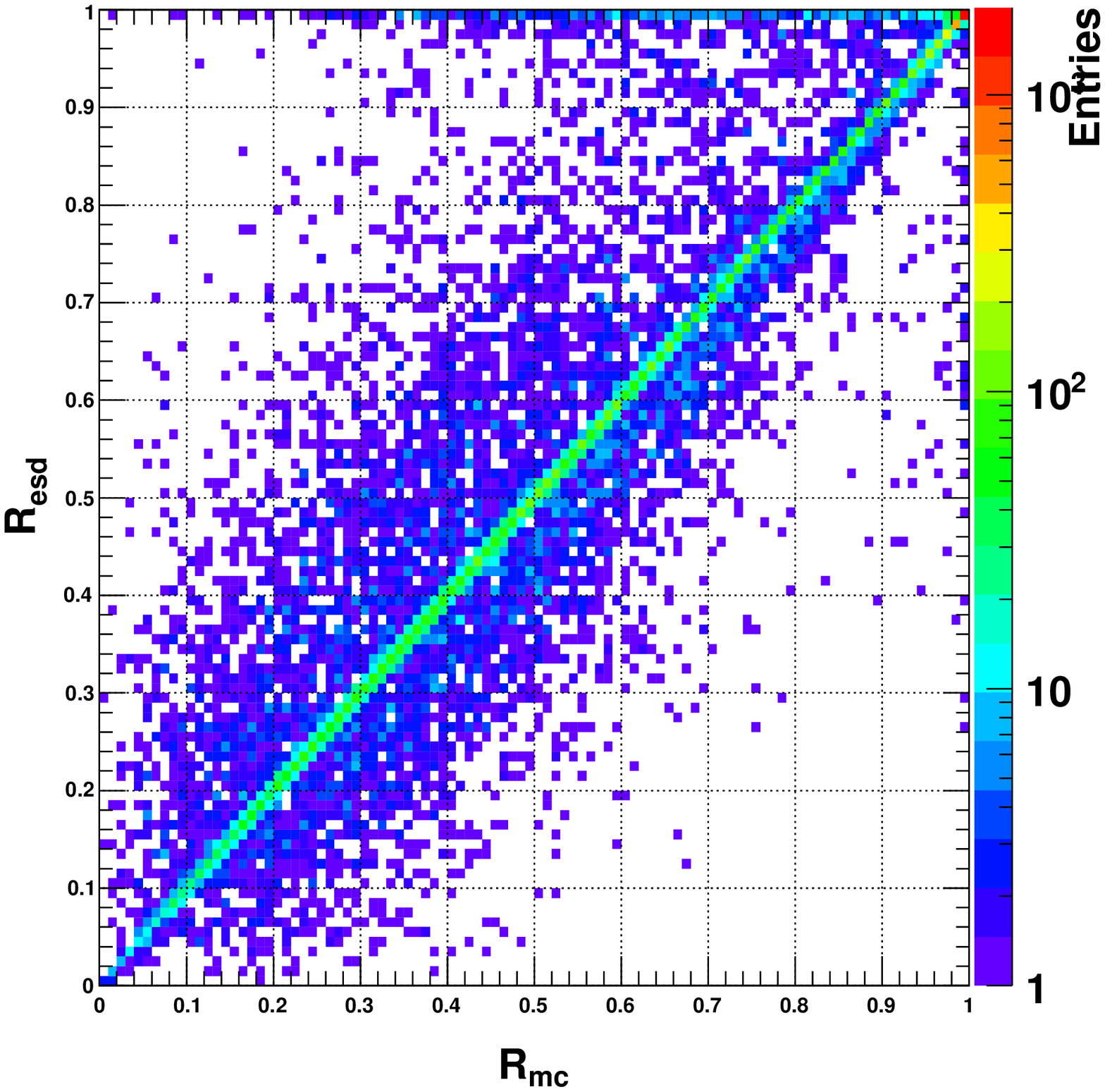} % Please
  }\vspace{0.5cm}
  \caption{Correlation between the quantity $R$ computed from MC information and from TPC+ITS information.} \label{Rjetjetmcesd}
  \end{center}
\end{figure}

  In the Fig. \ref{Rjetjetmcesd},
we show the analogous plot for the variable $R$.

For describing the topology of the events we performed an
azimuthal correlation. The idea is to select the leading particle,
and apply a rotation  placing it at $\pi/2$ rad. After that, we
plot the azimuthal distribution of the associated particles with
respect to the leading one. In the following our convention about
the azimuthal correlation it will be referred as $\Delta\phi$ and
it refers to: $\Delta (\phi - \pi/2)$.

\begin{figure}
\begin{center}
\resizebox{0.7\textwidth}{!}{ \includegraphics{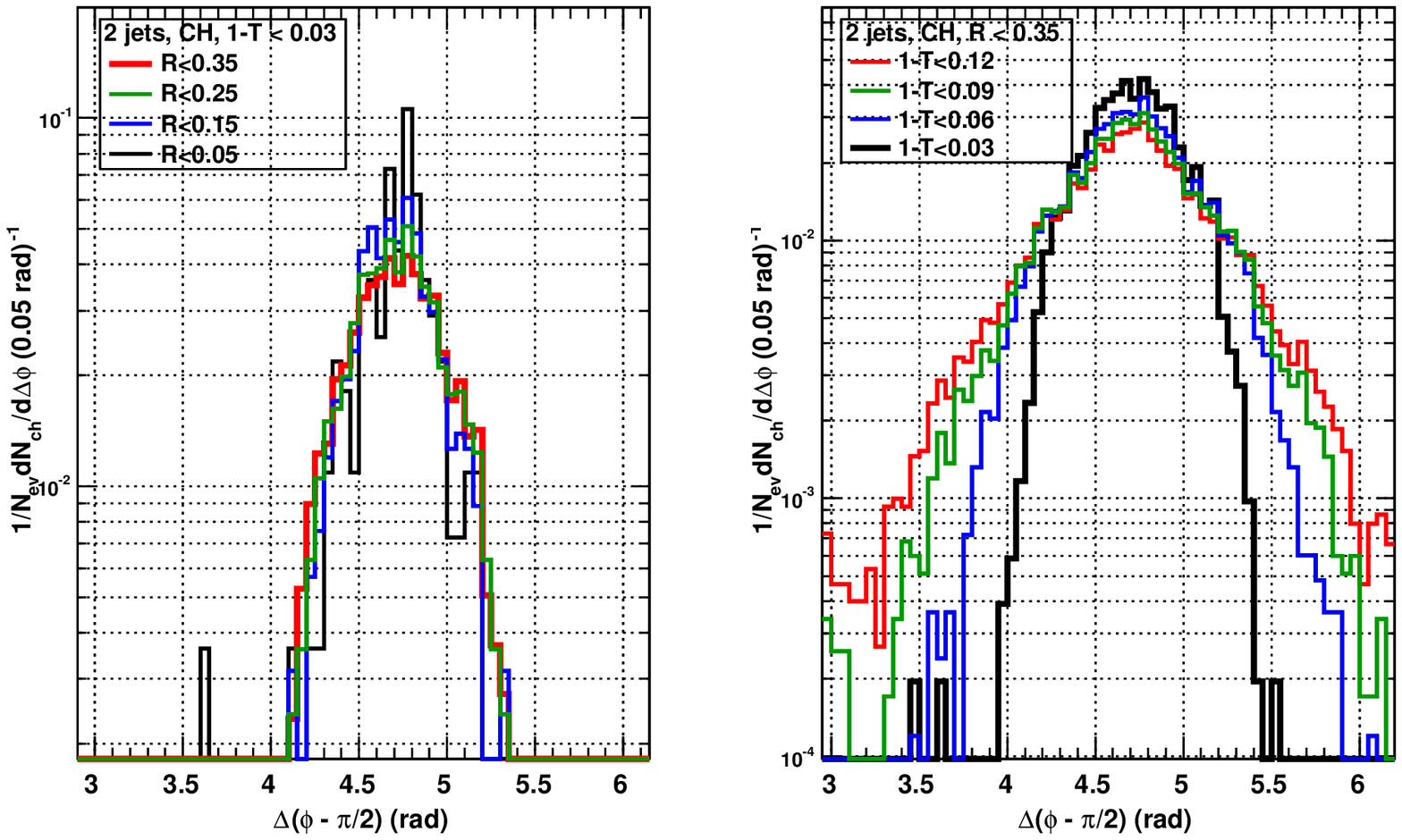} % Please
  }\vspace{0.5cm}
  \caption{Sensitivity of the azimuthal correlation of the away side peak on the thrust variable
  range. Fixed $1-T$ interval and variation of the $R$ range (right panel). Fixed $R$
  and modification of the range $1-T$ (left panel)} \label{efectoR}
  \end{center}
\end{figure}

If we concentrate ourselves to events in the region $1-T\leq0.03$,
and small $R\leq0.35$ values. The azimuthal correlation of the
Fig. \ref{efectoR} (right panel) shows that the width of the away
side peak does not change if we modify the cut in the $R$ range.
This is in agrement with our assumption which suggests that $R$ is
important to select complete events in the acceptance. On the
other hand, if we select events with small value of $R$
($R\leq0.35$) and modify the range of $1-T$ (Fig. \ref{efectoR}
(left panel)) there is a clear evolution in the structure of the
away side peak. If we increase the value of $1-T$ the selected
events include different configuration multijets (split jets)
which are manifested in the azimuthal distribution behavior. This
suggests that the cut $1-T\leq0.03$ is almost equivalent to select
particles inside a cone radii
$R_{c}=\sqrt{(\Delta\phi)^{2}+(\Delta\eta)^{2}}=1$.

 In the following we will concentrate on particular parts of the map, plotting the azimuthal correlations encountered.

 The
 procedure is:
\begin{itemize}
\item Select events according to their values of shape
 variables as shown in table \ref{tabla0}

 \begin{table}[!hbt]
\begin{center}
 \begin{tabular}{|c|c|c|}
 \hline
 Region & Kind of event & Variables  \\
 \hline \hline
 A & Dijets & $R\leq0.35$, $\tau\leq0.03$ \\
\hline
 B & Monojets & $R\geq0.9$, $\tau\leq0.03$ \\
 \hline
 C & Mercedes & $R\leq0.4$, $\tau\geq0.25$ \\
 \hline
 \end{tabular}
 \caption{1-T and R parameters used for the present analysis.}
 \label{tabla0}
\end{center}
\end{table}

\end{itemize}

\begin{figure}
\begin{center}
\resizebox{0.6\textwidth}{!}{ \includegraphics{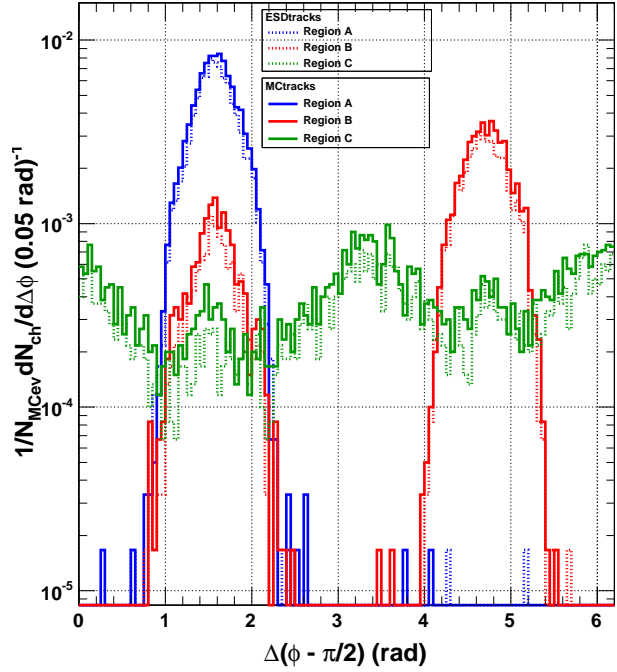} % Please
  }\vspace{0.5cm}
  \caption{Azimuthal correlation for particles with $p_{t}\geq1.5$ GeV/c: dijets (red-region A), monojets (blue-region B)
  and the mercedes events (green-region C). True spectrum (solid line), measured spectrum (dotted line).
  The leading particle is not shown.} \label{jetjetazimutal}
  \end{center}
\end{figure}

The azimuthal correlations of events sited at different regions of
the thrust map are plotted in Fig. \ref{jetjetazimutal}. The
evolution of the away peak (formed by particles with
$\pi\leq\Delta\phi\leq2\pi$) is interesting.

As predicted, the events of region B really have a monojet
topology in the azimuth. One can see there are associated
particles which go near to the leading particle (peak in the
toward side: $0\leq\Delta\phi\leq\pi$) but in the away side there
is no corresponding jet. In contrast, for events of the region A
the peak of the away side is located at $\Delta\phi\sim3\pi/2$,
so, in the transverse plane we have 2 back-to-back jets. In the
case of the events of region C we found a three-jet structure, in
the green distribution we observe three peaks in the spectrum: the
first (associated to the leading jet) at $\Delta\phi\sim\pi/2$ and
the others at $\Delta\phi\sim7\pi/6$ and $\Delta\phi\sim11\pi/6$
respectively. Due to the observed topology we named the latter
``mercedes'' events. The result bears some resemblance with the
away side structure observed at RHIC e.g. PHENIX collaboration in
heavy-ion collisions \cite{:2008cqb}. This observation brought us
to study the presence of the same
 double hump structure at RHIC energies \cite{Ayalaetal2}

\begin{figure}
\begin{center}
\resizebox{1\textwidth}{!}{ \includegraphics{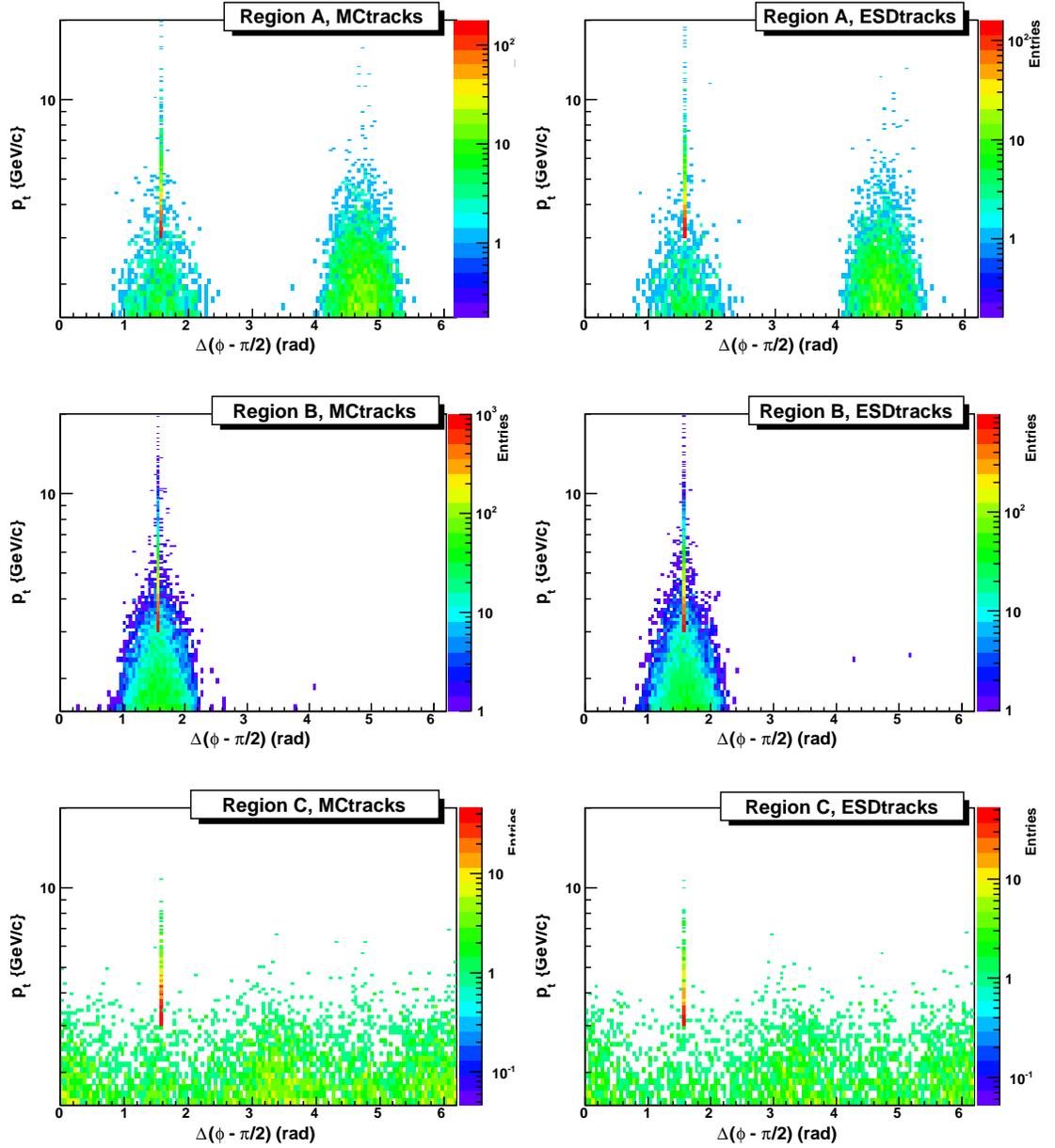} % Please
  }\vspace{0.5cm}
  \caption{Transverse momentum vs. $\Delta (\phi - \pi/2)$ for the associated particles: dijets (top), monojets (middle)
  and the mercedes events (bottom). True distribution (left plots), measured distribution (right plots). The leading particle is shown,
  and you can note the cut $p_{t}\geq1.5$ GeV/c which we imposed.}
\label{ptvsphi}
 \end{center}
\end{figure}
%\newpage

In the Fig. \ref{ptvsphi} we show for the different regions of the
thrust map, a two dimensional distribution: $p_{t}$ vs. $\Delta
\phi$ for the associated particles.

The table \ref{tabla1} shows a summary of the analyzed events.

 \begin{table}[!hbt]
\begin{center}
 \begin{tabular}{|c|c|c|c|c|c|}
 \hline
 Event & MCtracks & ESDtracks & Efficiency & $1-T$ ($\tau$) cuts & $R$ cuts \\
 \hline
\hline
 With $T$ & 28920 & 24960 & $86\%$& no & no \\
 \hline
 Dijet & 1503 & 1316 &67$\%$ &$\tau\leq0.03$ & $R\leq0.35$ \\
 \hline
 Monojet & 8903 & 8329 & 79$\%$ &$\tau\leq0.03$ & $R\geq0.9$ \\
 \hline
 Mercedes & 523 & 439  & 62$\%$ &$\tau\geq0.25$ &$R\leq0.4$ \\
 \hline
 \end{tabular}
 \caption{Number of events with particular topologies in the analyzed sample of 1.2 million of events.}
 \label{tabla1}
\end{center}
\end{table}
The efficiency represents the number of reconstructed events of
given topology with respect to the generated events.

One observes a $\sim70\%$ efficiency with respect to the MC. This
may indicate that due to reconstruction and$/$or absorption of
particles one relocates events in other parts of the thrust map as
shown in Figs. \ref{thrustjetjetMCESD}-\ref{Rjetjetmcesd}.

Note that, out of the 1200000 events only $2.4\%$ (with given $T$)
of them pass the cuts imposed (at least 1 particle with
$p_{t}\geq3$ GeV/c and $|\eta^{leading}|\leq0.5$). The dijets in
acceptance reach about $\sim 0.125\%$ of the total, and about
$0.04\%$ belong to mercedes event types. About $37.8\%$ of the
accepted events belong to clearly identifiable categories, while
the others correspond to events with multi-jets closer to the
leading jet.

In the next section we will study each one of the topologies
selected. That study includes a visualization of the events
through the use of the tools of AliRoot.\footnote{AliRoot is the
name of ALICE Off-line framework for simulation, reconstruction
and analysis. It uses the ROOT system as a foundation on which the
framework and all applications are built.}

%\newpage
\subsection{Dijets and monojets}
\begin{figure}
\begin{center}
\resizebox{0.8\textwidth}{!}{ \includegraphics{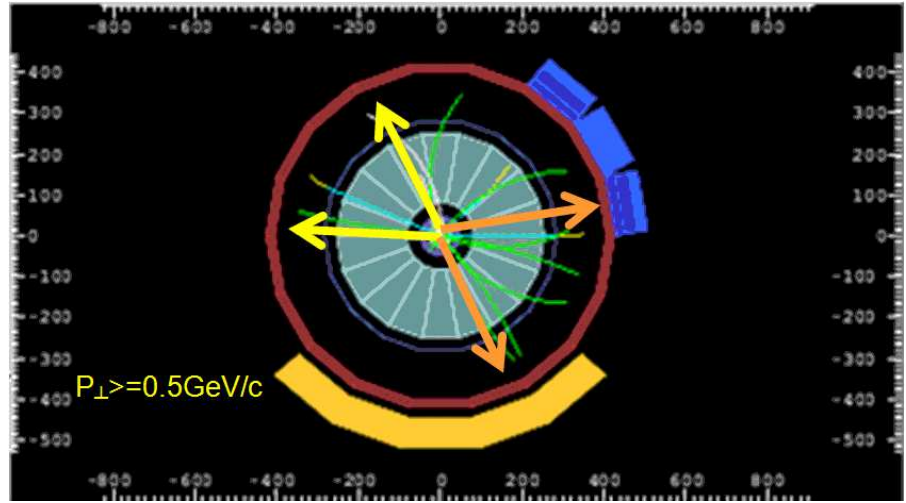} % Please
  }%\vspace{0.5cm}
  \caption{Visualization of one dijet event from the region A.   The lines are
  primary monte carlo tracks with $p_{t}\geq0.5$ GeV/c. The superimposed arrows separate each jet for a better visualization.} \label{dijet1550}
 \end{center}
\end{figure}

In the present section we are investigating the $p_{t}$ spectra
and the multiplicities of the jets in various configurations. As a
first step we turn to the visualization tool of events in ALICE.
We selected events located in the region A and scanned them. Due
to their small $R$ values, they should be inside the acceptance.
Fig. \ref{dijet1550} shows the visualization of one of them. This
particular event has $R=0.00132$ and $1-T=0.0000896$. The lines
are primary monte carlo tracks with $p_{t}\geq0.5$ GeV/c. The
figure clearly shows the whole event contained inside the TPC.

Looking at more events shows always the same structure in the
visualization.

Further we computed the total transverse momenta in each jet (here
we used all events of the region A).
\begin{figure}
\begin{center}
\resizebox{0.5\textwidth}{!}{ \includegraphics{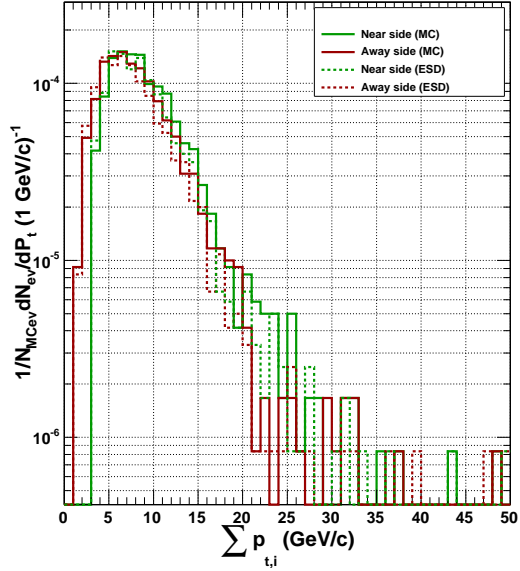} % Please
  }%\vspace{0.5cm}
  \caption{Distribution of the sum of the transverse momenta of particles with $p_{t} > 0.3$ GeV/c in the toward and away regions for dijet events. The toward region
  (black, leading jet) is formed by primary charged particles in the azimuthal range: $\pi/4$ rad $\leq \Delta \phi \leq$ $3\pi/4$ rad while the
  away region (red line) corresponds to particles with: $5\pi/4$ rad $\leq
\Delta \phi \leq$ $7\pi/4$ rad. The plot shows two cases: reconstruction
(dotted line) and generation (solid line).} \label{ptconservdijet}
 \end{center}
\end{figure}
In order to do it, we divided the event into two parts: toward and
away. The first (near side) contains the leading particle and it
is formed by all particles in the interval: $\pi/4$ rad $\leq
\Delta \phi \leq$ $3\pi/4$ rad. The away side is formed by
particles in the interval: $5\pi/4$ rad $\leq \Delta \phi \leq$
$7\pi/4$.

\begin{figure}
\begin{center}
\resizebox{0.8\textwidth}{!}{ \includegraphics{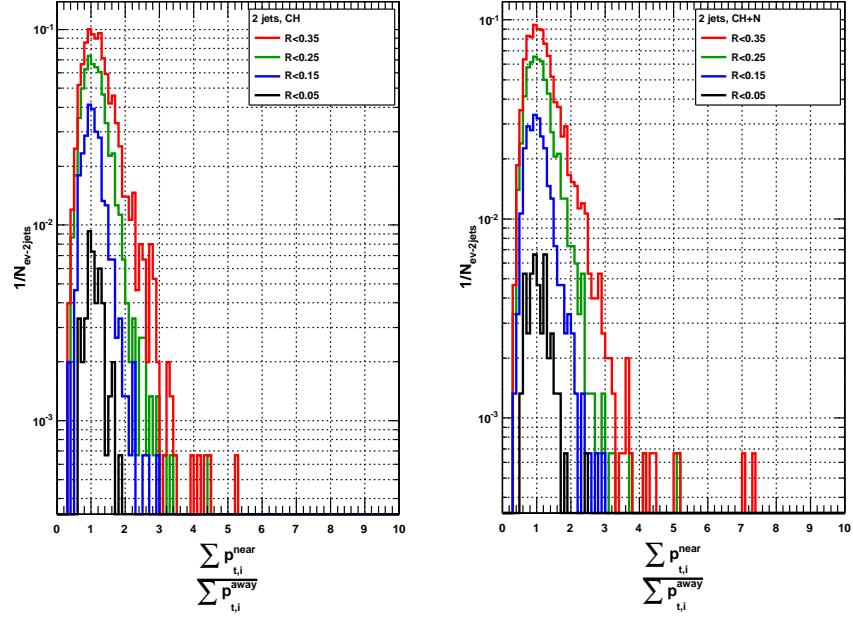} % Please
  }%\vspace{0.5cm}
  \caption{Ratio of the transverse momenta of the toward jet over transverse momenta of the away jet as a function of $R$ for dijet events.
  The participants have $p_{t}\geq0.3$
  GeV/c. The near side corresponds to the azimuthal range: $\pi/4$ rad $\leq \Delta \phi \leq$ $3\pi/4$ rad.
   For the away side: $5\pi/4$ rad $\leq\Delta \phi \leq$
   $7\pi/4$. Charged component (right), (left) including the
   neutral component.
 } \label{ptratiodijet}
 \end{center}
\end{figure}

 The Fig. \ref{ptconservdijet} shows
the distribution of the transverse momentum spectrum of each jet
(sum in each azimuthal region of the transverse momentum of all
the participants with  $p_{t}\geq0.3$ GeV/c) for generation and
reconstruction. The away side distribution is left shifted $\sim1$
GeV/c with respect to the near side one. This can be understood in
terms of the fluctuation of the neutral component of the
associated jet and also as effects from the acceptance in the
associated jet. In order to illustrate this arguments we plotted
the distribution of the ratio: total transverse momenta in the
``toward'' side over the total transverse momenta in the ``away''
side for the generation case. The Fig. \ref{ptratiodijet} (right
panel) shows the behavior of such distribution (red line). The
distribution manifest a clear peak at 1, this fact is in agrement
with our assumption about the dijet structure. However the
distribution shows many events where the near side jet represent
up to 5 times the energy of the away side jet. In this respect the
role of the value $R$ is important, because as you can see the
width of the distribution decreases as the $R$ interval is
decreased. The same analysis can be performed including the
neutral component (left panel). The away side jets with a low
transverse momentum correspond to events where the away jet is not
completely contained.

If you refer to Appendix A of this note, you can convince about
ESA allows rejecting events that some jet finders as JETAN could
reconstruct as a perfect di-jet, although part of the jet stays
out of the acceptance.

\begin{figure}
\begin{center}
\resizebox{0.5\textwidth}{!}{ \includegraphics{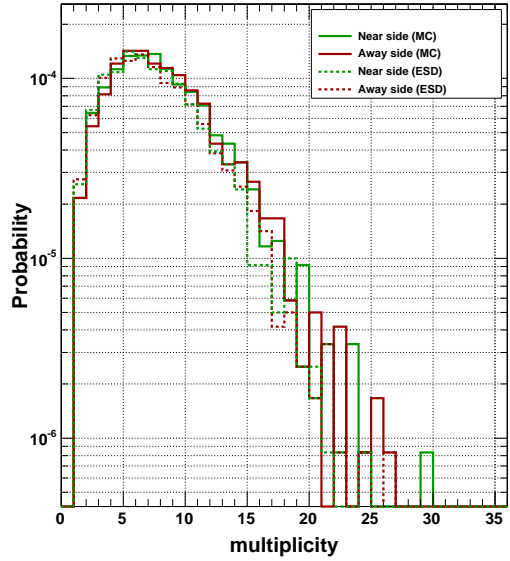} % Please
  }\vspace{0.5cm}
  \caption{Multiplicity distribution of particles with transverse momenta: $p_{t} > 0.3$ GeV/c for dijet events. The particles belonging to the
  leading jet (black line)
 consist of primary charged particles in the azimuthal range: $\pi/4$ rad $\leq \Delta \phi \leq$ $3\pi/4$ rad. The
  away side (red line) corresponds to particles with: $5\pi/4$ rad $\leq
\Delta \phi \leq$ $7\pi/4$ rad. The plot shows two cases:
reconstruction (dotted line) and generation (solid line).}
\label{mult2jet}
 \end{center}
\end{figure}

We checked also the multiplicity distributions for the dijets
events.  In the Fig. \ref{mult2jet}, we show the multiplicity
distribution for the away and toward sides.

\begin{figure}
\begin{center}
\resizebox{0.8\textwidth}{!}{ \includegraphics{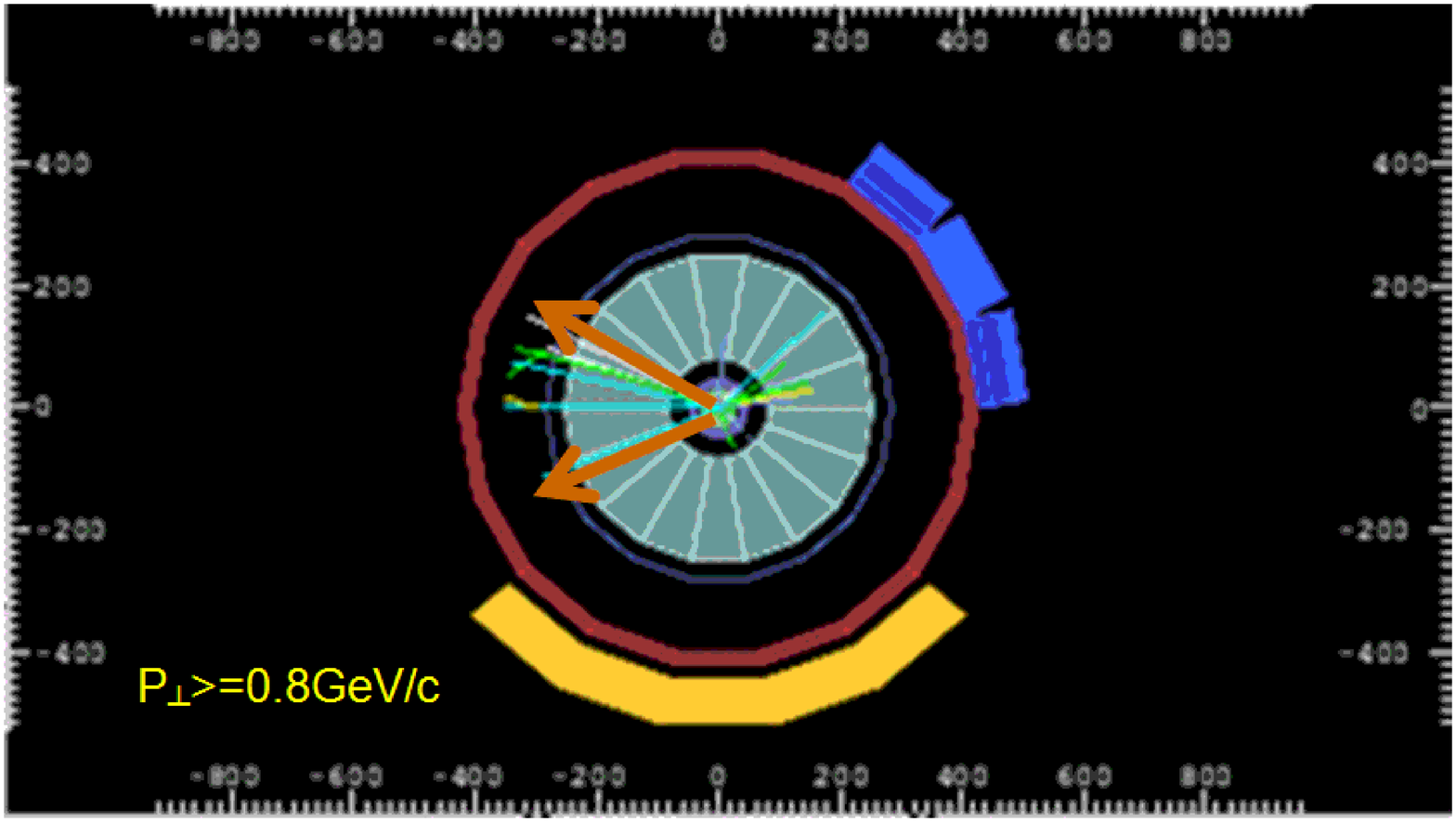} % Please
  }\vspace{0.5cm}
  \caption{Visualization of a mono-jet event.
  The lines are
 mctracks with $p_{t}\geq0.8$ GeV/c. The arrows contain the particles associated with the jet.} \label{vismono}
 \end{center}
\end{figure}

A  mono-jet event taken from region B of the ESA map is shown  in the Fig. \ref{vismono}.

\begin{figure}
\begin{center}
\resizebox{0.5\textwidth}{!}{ \includegraphics{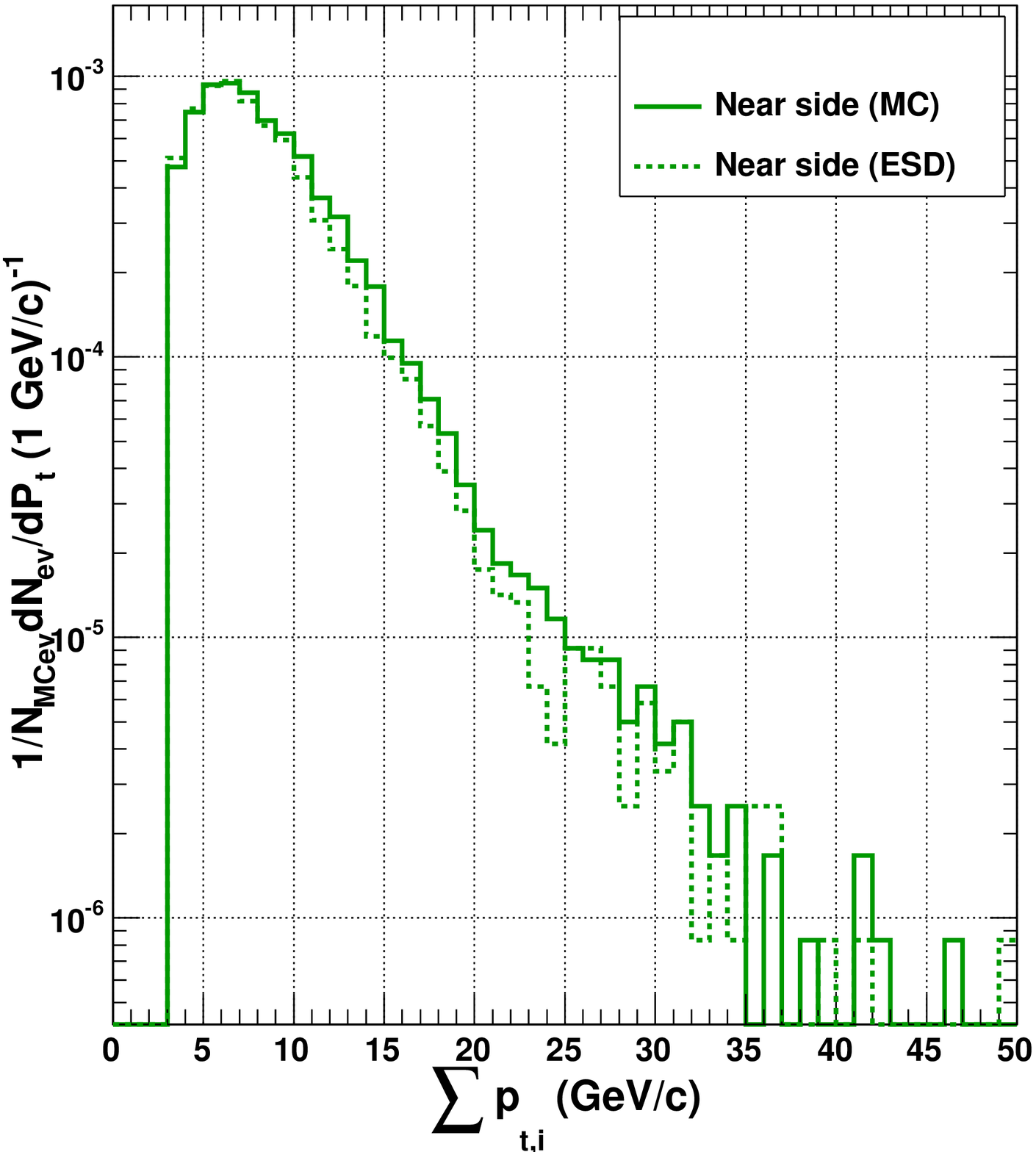} % Please
  }%\vspace{0.1cm}
  \caption{The distribution of the total charged transverse momenta of the identified mono-jets.
  The particles which were counted are within the azimuthal range: $\pi/4$ rad $\leq \Delta
\phi \leq$ $3\pi/4$ rad; and they have: $p_{t}>0.3$ GeV/c. The
generated (solid line) and the reconstructed (dotted line) are
shown.} \label{monopt}
 \end{center}
\end{figure}

The lines correspond to mctracks with $p_{T}\geq0.8$ GeV/c. For
this event: $1-T=0.00216$ and $R=0.99874$.
  Again, by counting primary charged
particles within the azimuthal range: $\pi/4$ rad $\leq \Delta
\phi \leq$ $3\pi/4$ rad, we estimate the total transverse momentum
of the jet. In the Fig. \ref{monopt}, is the distribution of the
total transverse momenta for events of region B. Note that the
peak of this distribution is at $\sim 6$ GeV/c, in the case of
di-jet events this peak is also located at $\sim 6$ GeV/c.
% As we discussed previously, this fact imply that the reconstruction in TPC looks wrong for this class of events. Note also, that in both cases the identified jets reach $\sim 70$ GeV/c of charged transverse momenta.

\begin{figure}
\begin{center}
\resizebox{0.5\textwidth}{!}{ \includegraphics{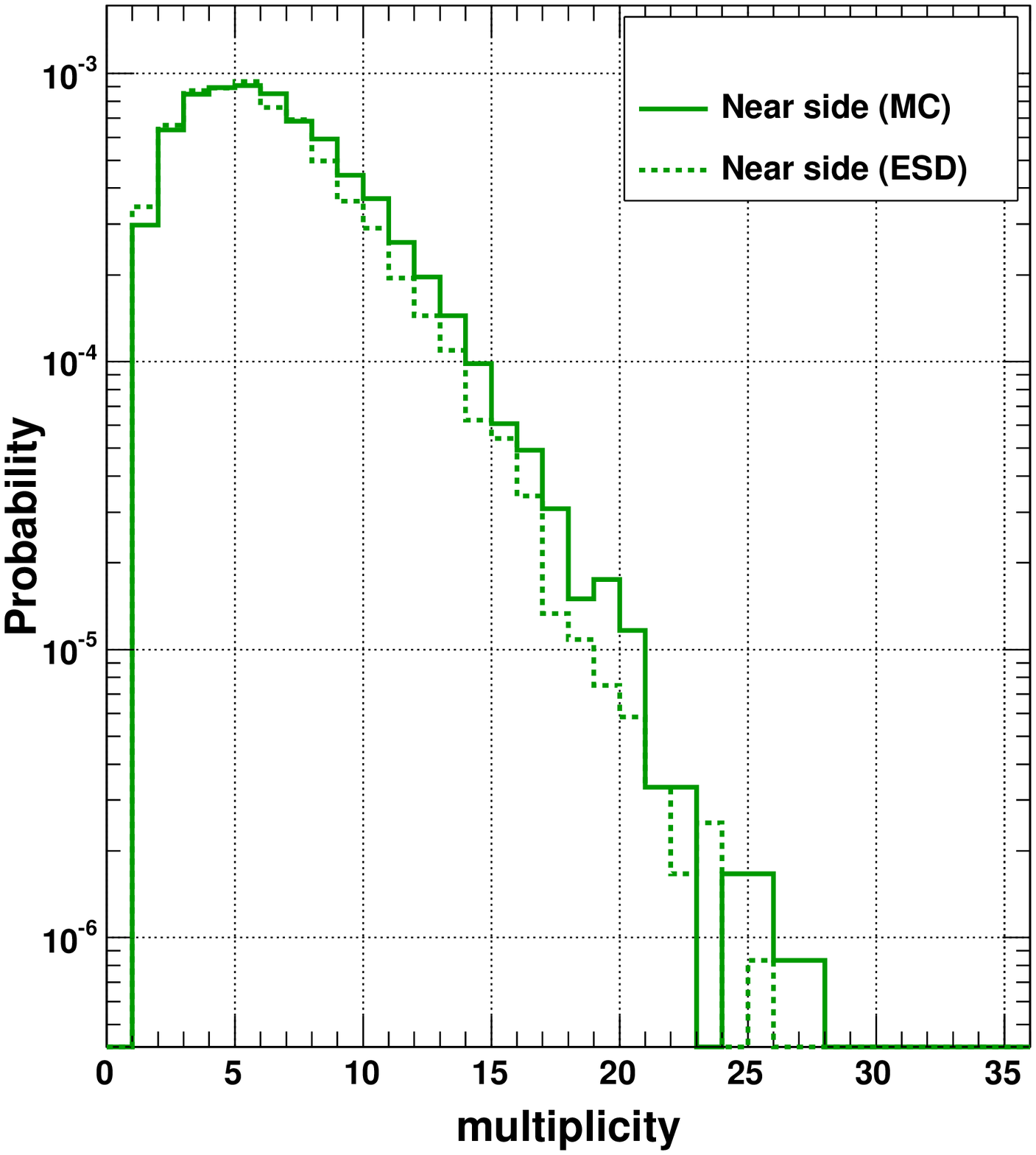}}
  \caption{Multiplicity of primary charged particles ($p_{t}>0.3$ GeV/c) of mono-jet events. The azimuthal range is: $\pi/4$ rad $\leq \Delta
\phi \leq$ $3\pi/4$ rad. The results of generation (solid line)
and the reconstruction (dotted line) are compared.}
\label{multmono}
 \end{center}
\end{figure}

The multiplicity distribution for mono-jet events is in Fig.
\ref{multmono}.

The conclusion of this part of the analysis is, that according
with the results of the visualization and the behavior of the
multiplicity and transverse momentum spectra ESA
 works fine for discriminating the di-jet events from the mono-jet ones.   Using the event shape analysis the
signals can be cleaned in order to improve the jet studies.

\vspace{0.5cm}
 \vspace{0.5cm}

\subsection{Three-jet events}
\vspace{0.5cm}
 The green distributions of the Fig.
\ref{jetjetazimutal} allows to observe a double hump structure in
the away side of the azimuthal distribution. They look like if the
three particles with the highest $p_{t}$ in each event were
distributed in the transverse plane according to: the leading
particle at $\pi/2$ radians, and the others at: $7\pi/6$ radians
and $11\pi/6$ radians, respectively.

\begin{figure}
\begin{center}
\resizebox{0.9\textwidth}{!}{ \includegraphics{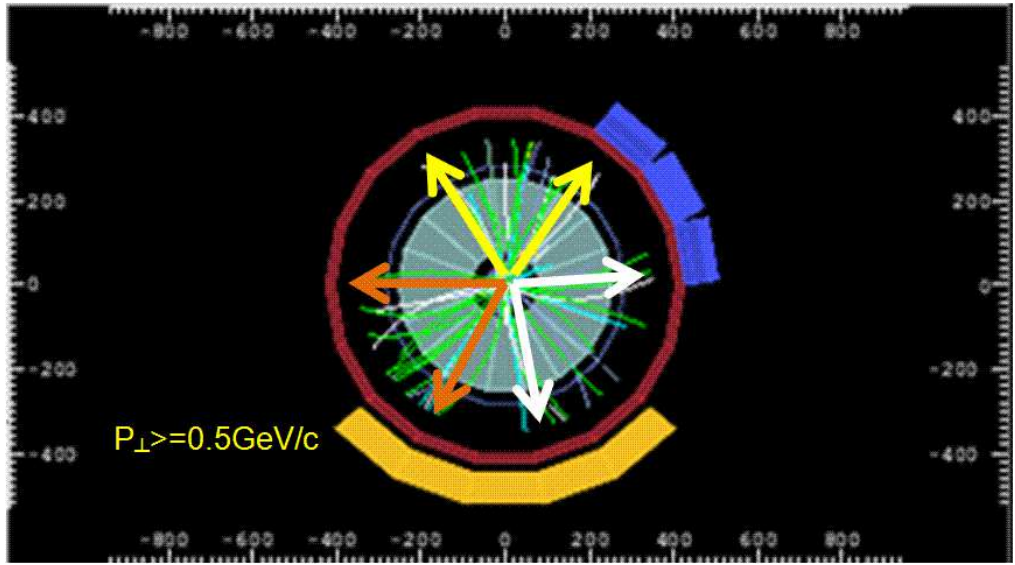} % Please
  }%\vspace{0.5cm}
  \caption{Scan of one event of region C.  Particles which appear
  in the picture have: $p_{t}\geq0.5$ GeV/c.} \label{visthreejet}
  \end{center}
\end{figure}

There are many configurations of isotropic distributions that
could lead to high $1-T$ values,
 however the scans of the events as shown in Fig. \ref{visthreejet} for a typical event result in a clear three jets configurations.
 A small contribution from events with more than 3 jets can be
 found.

It is important to say that this class of events occurs completely
inside the acceptance of our detector, we have to remind that this
is controlled by the term $R$. For example, if we increase the
range of $R$, the azimuthal distribution in the away side shows a
shift of the two peaks because the calculation of the variables
use incomplete information of the event.

\begin{figure}
\begin{center}
\resizebox{0.5\textwidth}{!}{ \includegraphics{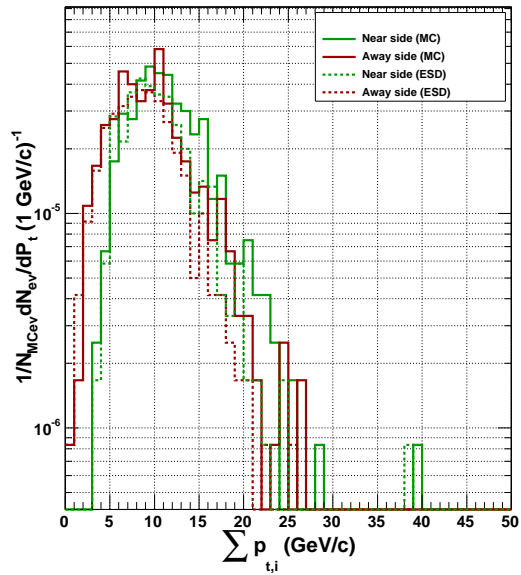} % Please
  }%\vspace{0.5cm}
  \caption{Transverse momentum spectra for three-jet events, the including charged particles with $p_{t}>0.3$ GeV/c. Toward side
  (black line) corresponds to particles in the azimuthal region: $\pi/4$ rad $\leq \Delta \phi \leq$ $3\pi/4$ rad. Away side
  formed by particles in the remainder azimuth.   The results at generator
level (solid line) and reconstruction (dotted line) are shown.}
\label{pttresjets}
  \end{center}
\end{figure}

In order to see the conservation of the transverse momentum in
this class of events, we divided the azimuth into three regions:
\begin{itemize}
\item[] \textbf{Near side:} formed by particles in the azimuthal
range: $\pi/4$ rad $\leq \Delta \phi \leq$ $3\pi/4$ rad.  \item[]
\textbf{Away side:} formed by the particles in the remainder of
the azimuth.
\end{itemize}
As in the previous cases we have taken into account only primary
charged particles with $p_{t}>0.3$ GeV/c. The Fig.
\ref{pttresjets} shows the transverse momentum of each side. The
agreement between both spectra is reasonable in the limits of the
statistics.

\begin{figure}
\begin{center}
\resizebox{0.8\textwidth}{!}{ \includegraphics{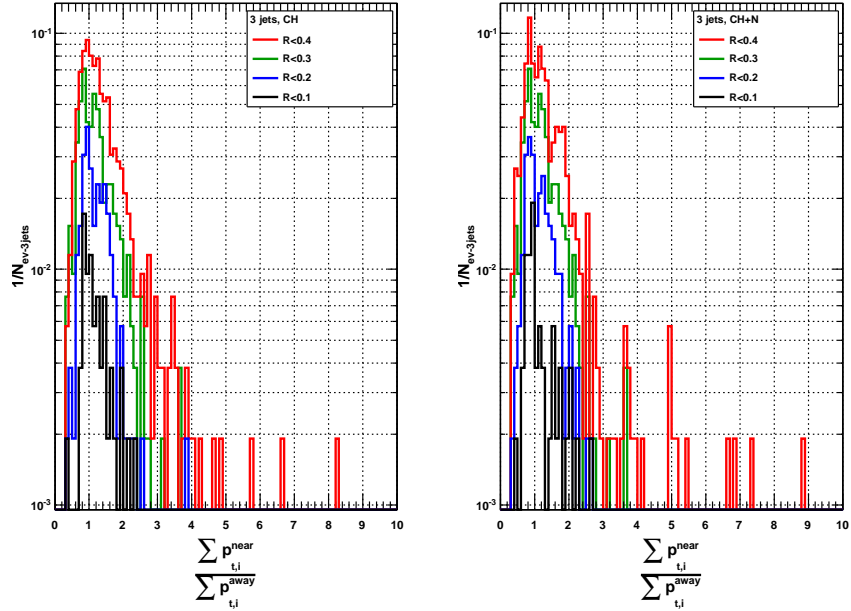} % Please
  }%\vspace{0.5cm}
  \caption{Mercedes events. Ratio of transverse momentum distributions of toward jet over transverse momenta of the away jets. The
   $\Delta \phi$ ranges used are the same as in the previos figure.
 } \label{ratio3jetjetjet1550}
  \end{center}
\end{figure}

We compute also event by event the ratio of the transverse momenta
of the toward jet over the vectorial sum of the away jets as a
function of $R$. The result of this analysis is shown in the Fig.
\ref{ratio3jetjetjet1550}. Again these distributions reach their
peaks at 1, suggesting the correct transverse momentum
conservation.

The events with mercedes topology were found in MB simulation at
10 TeV in the c. m. as well as in MB simulations at
200 GeV \cite{Ayalaetal2}.

%\newpage

\newpage
\subsection{Multiplicity in the context of ESA.}

\begin{figure}
\begin{center}
\resizebox{0.8\textwidth}{!}{ \includegraphics{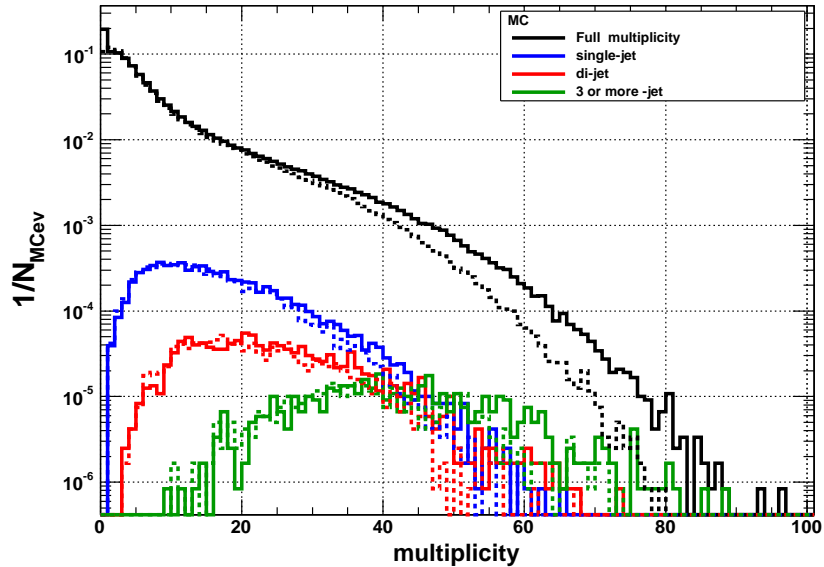} % Please
  }%\vspace{0.5cm}
  \caption{Multiplicity (charged paricles with $p_{t}>0.3$ GeV/c) distribution. In black   is the full true spectrum.
  The different colored distributions correspond to the three classes of topologies which we discriminated using
  ESA.}
\label{multiplicityfull}
 \end{center}
\end{figure}

In this section we investigate the multiplicity of each event
which we selected.  In order to do this task we plotted the
multiplicity spectra of the full sample (1 200 000 events), and we
compared it with the multiplicities of the events with a given
thrust value. The multiplicity is the number of primary charged
particles in the acceptance $|\eta|\leq1$ with $p_{t}>0.3$ GeV/c.
As we can see in the plots of the Fig. \ref{multiplicityfull}, the
conditions which we demanded to each event reduces the number of
low multiplicity events. The events with are not related with any
of the classes we studied belong to multi-jet events, and also are
of high multiplicity. We observe that the mercedes events are
generally of large multiplicities.

\section{Bulk analysis}
\begin{figure}
\begin{center}
\resizebox{1\textwidth}{!}{ \includegraphics{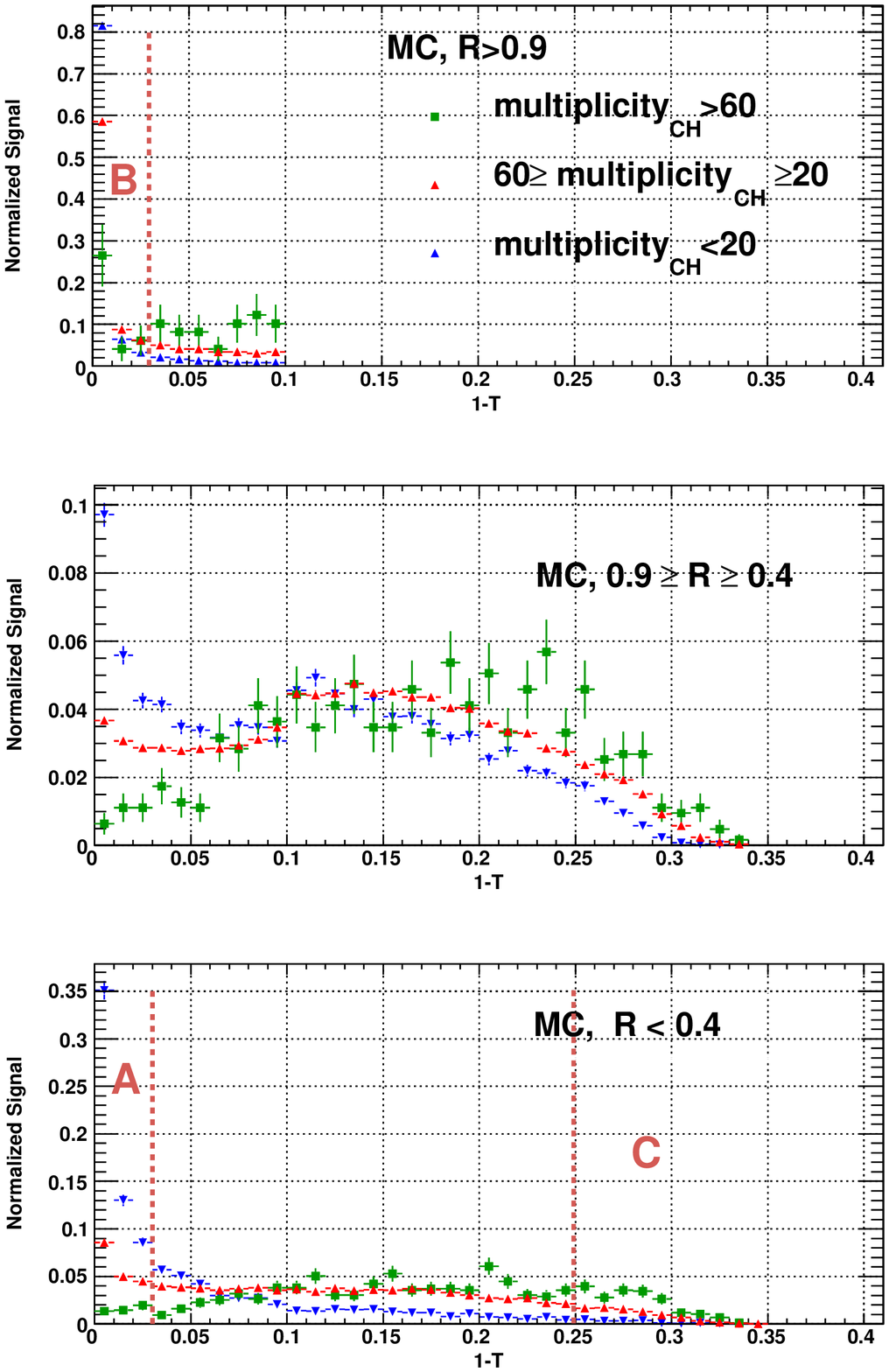} % Please
  }%\vspace{0.5cm}
  \caption{Projection on the 1-T axis as a function of R and multiplicity. The regions with interesting topologies are indicated.}
\label{multiplicityvsTandR}
 \end{center}
\end{figure}

The analysis of the event shape space projected on the $1-T$
offers also interesting results. Without detailed analysis of tje
jet components one may directly compare with the different
generators. In the following we present the results of each
analysis for the generators: Pythia and Phojet. Another
interesting question is what happens with the events which thrust
and recoil values are outside of the regions A, B and C. In the
figure \ref{multiplicityvsTandR} there is the $1-T$ normalized
spectrum for different multiplicity bins. Three histograms are
shown, the upper one corresponds to the cut $R<0.9$, the middle:
$0.9\geq R \geq0.4$; and finally the events with $R<0.4$ appears
in the bottom one. It is clear that the distributions are quite
different in the cases $R>0.9$ and $R \leq0.4$. In the last one
the probability of finding a high multiplicity mercedes event is
bigger than the probability of finding a low multiplicity event
with mercedes topology. For events of region A, the maximum of the
distribution is attained for the lowest multiplicity events. Note
also, that in the middle $R$ region (unexplored yet); the
spectrums associated to different multiplicity bins are quite
similar in their shape.

%\newpage
\subsection{Comparison between Pythia and Phojet in the context of
ESA}

The Monte Carlo generators Phojet \cite{phojettessis} and Pythia
\cite{pythia} use both LO QCD matrix elements for the hard
scattering sub-processes. Initial and final state parton radiation
and the string fragmentation model are included as implemented in
the JETSET program \cite{pythia2}. The two Monte Carlo generators
differ in the treatment of multiple interactions and the
transition from hard to soft processes at low transverse parton
momentum. The hard parton-parton cross-section diverges towards
low transverse momenta and therefore needs a regularization to
normalize to the measured total cross-section.

 Hadronic collisions at high energies involve
the production of particles with low transverse momenta, the
so-called soft multi-particle production. The theoretical tools
available at present are not sufficient to understand this feature
from QCD alone and phenomenological models are typically applied
in addition to perturbative QCD. The Dual Parton Model (DPM)
\cite{dpm}  is such a phenomenological model and its fundamental
ideas are presently the basis of many of the Monte Carlo
implementations of soft interactions.

 The Monte Carlo event generator Phojet can be used to simulate
hadronic multi-particle production at high energies for
hadron-hadron, photon-hadron, and photon-photon interactions with
energies greater than 5 GeV. It implements the DPM as a
two-component model using Reggeon theory for soft and leading
order pQCD for hard interactions. Each Phojet collision includes
multiple hard and multiple soft pomeron exchanges, as well as
initial and final state radiation. In Phojet pQCD interactions are
referred to as hard Pomeron exchange. In addition to the model
features as described in detail in \cite{phojettessis}, the
version 1.12 incorporates a model for high-mass diffraction
dissociation including multiple jet production and recursive
insertions of enhanced pomeron graphs (triple-, loop- and
double-pomeron graphs).

So, Phojet provides an alternative to Pythia for the study of
processes that cannot be calculated with pQCD, such as minimum
bias events (events with high cross section and low transverse
momentum) and the underlying event activity in events with a high
transverse momentum parton-parton collision.

\subsection{Implementation of ESA}
The Pythia and Phojet events used are part of the standard
simulations in ALICE. The samples are staged on alicecaf and they
correspond to proton-proton collisions at 10 TeV in the c. m.,
minimum bias, magnetic field of 0.5 T. This analysis was done
using 200 000 events in each sample.

\begin{figure}
\begin{center}
\resizebox{0.8\textwidth}{!}{ \includegraphics{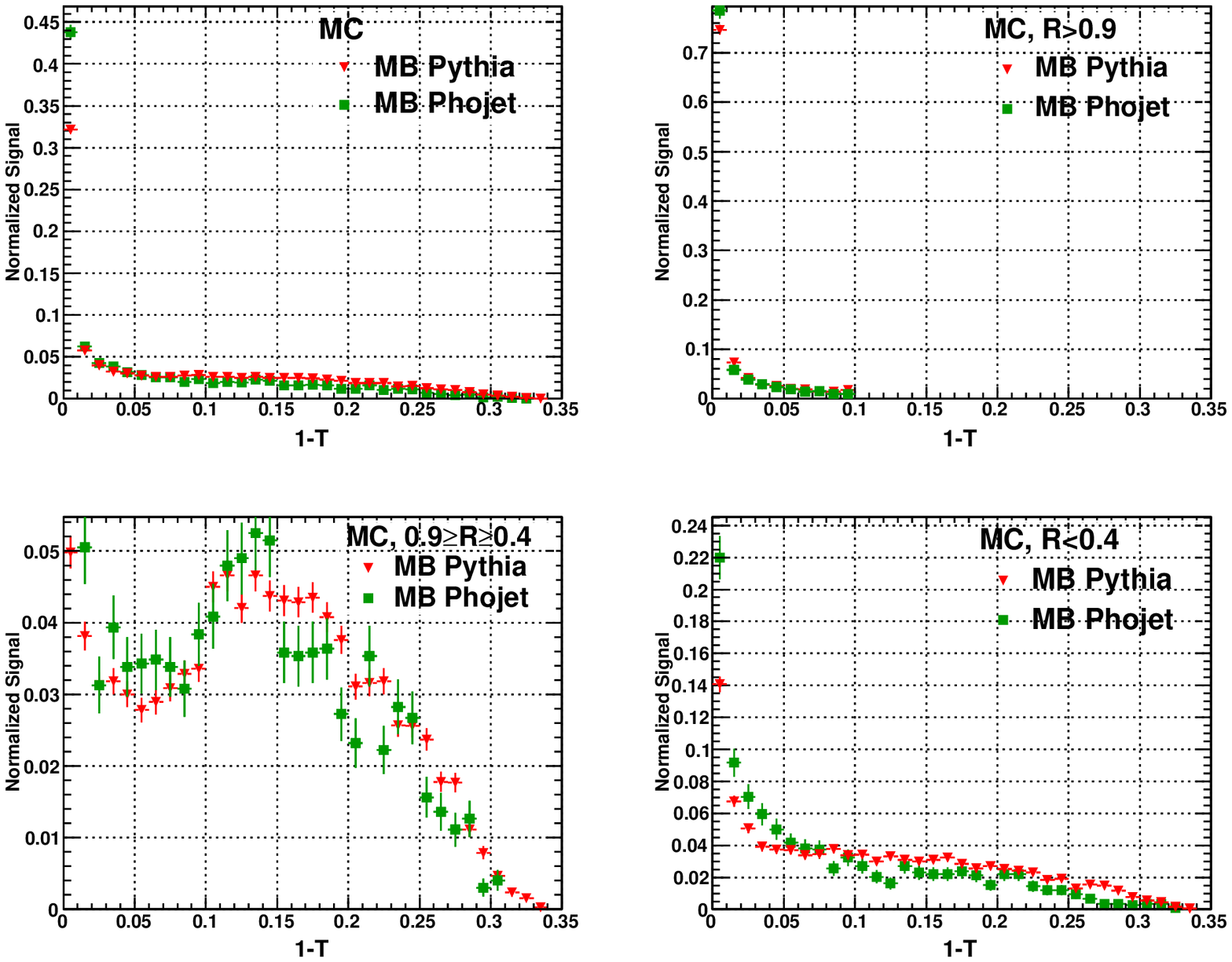} % Please
  }%\vspace{0.5cm}
  \caption{Projection on the 1-T axis for Pythia and Phojet generated events.}
\label{dif}
 \end{center}
\end{figure}

The Fig. \ref{dif} shows the comparison between the normalized
spectrum $1-T$ for the samples generated with Pythia and Phojet.
Three different regions in $R$ are explored: the first one:
populated by single-jet events ($R>0.9$). The second is the
intermediate $R$ zone ($0.4 \leq R \leq 0.9$). And finally the
zone where there are di-jet and three-jet events ($R<0.4$).

Note that in agrement with the discussion of the previous section,
we can observe in the region of the highest values of $R$ and
$1-T$ a more copious presence of mono-jet events in the
simulations generated by Phojet than in the case of Pythia ones.
The same situation appear in the case of di-jet events. Also, note
that the particles generated by Pythia look distributed in a way
more isotropic compared to the Phojet sample. For the more
isotropic topologies and especially
 the mercedes events the production with Pythia is more copious than with Phojet.

%\newpage
\section{Conclusions}
In the present note we have demonstrated the applicability of the
Event Structure Analysis in the case of measurements of charged
particle tracks in a detector of limited acceptance like ALICE.
The phase space in recoil and thrust variables offers a wide
variety of possible uses:
\begin{itemize}
\item[] Selection of well characterized events like monojets,
dijets and three jets ones. ESA is in that instance not a
replacement for any kind of jet finder but more of a preselection
that can be then studied with jet algorithms. As we report the
number of ``clean'' dijets for instance is minute in comparison
with the total sample. The use of conventional jet finding
algorithm in events with many jets as is the case at LHC leads
inexorably to the use of cuts that are sometimes leading to
results difficult to interpret. \item[]Bulk comparison of the
event characteristics with existing models. Since in the majority
of cases we are confronted with multijets events we believe useful
to establish a way to compare model predictions with projection on
the sphericity axis of events belonging to different recoil and/or
multiplicity intervals. \item[]Last but not least any deviation of
the present predictions of generator would lead to a rapid
detection of events located in an ``odd'' region of the $R$ vs.
$(1-T)$ phase space.
\end{itemize}

We have presented the discrimination power of ESA for specific
topologies: dijets events, monojets and three jet topologies.
According to the results presented, the ESA may be of use in the
physics analysis starting with the first data since with a sample
of 400000 events of minimum bias events, there are many events
with the topologies which we discussed.
\section {Acknowledgement}
The authors would like to thank Dr. J-P Revol for suggesting the
present work and for judicious comments to the results; the
discussions with E. Cuautle and I. Dominguez were of help in
elucidating some aspects of the analysis presented here. One of us
(A.O.) would like to acknowledge the HELEN fellowships which
greatly helped in getting full proficiency in the use of the
ALIROOT framework. The work was performed under the projet
IN115808 and Conacyt P79764-F.

\newpage

\appendix{}
\section{Appendix}
\subsection{About the usefulness of ESA in combination with JETAN}

\begin{figure}
\begin{center}
\resizebox{0.7\textwidth}{!}{ \includegraphics{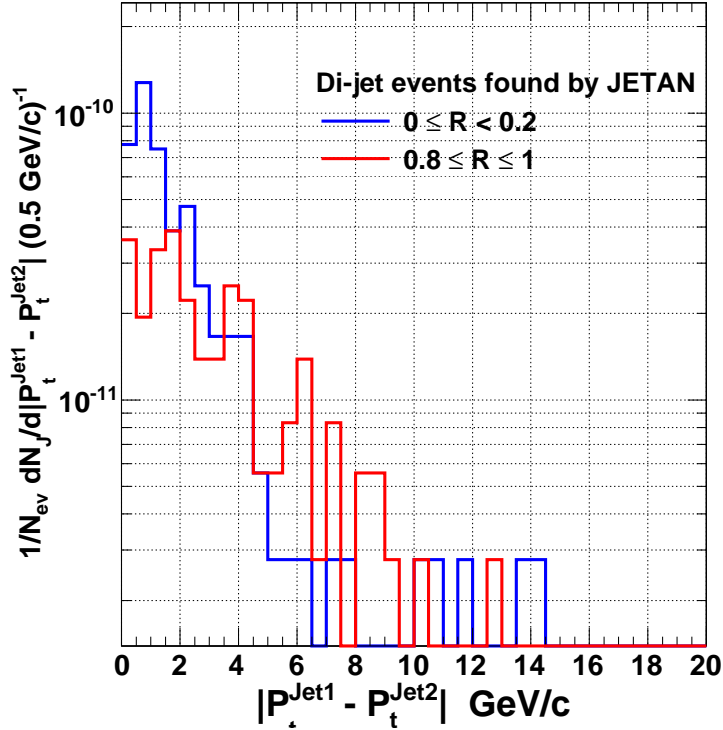} % Please
  }%\vspace{0.5cm}
  \caption{Analysis of the slice $1-T \leq 0.05$ of the thrust map.
  The jet finder UA1 was used for reconstructing the jets.
 } \label{jetan1}
 \end{center}
\end{figure}

\begin{figure}
\begin{center}
\resizebox{0.7\textwidth}{!}{ \includegraphics{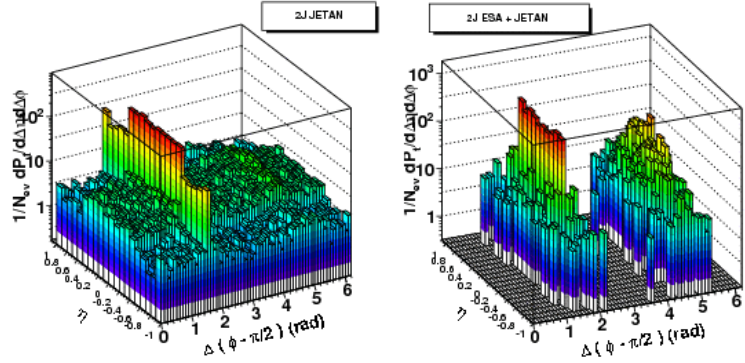} % Please
  }%\vspace{0.5cm}
 \caption{Topologies of the events identified as dijets. The transverse momentum distributions of the primary tracks in
  the plane $\eta$ vs $\Delta\phi$ (the leading particle is at
  $\Delta\phi=\pi/2$) for different samples of events are shown for the following cases.  Dijets reconstructed with
  JETAN (left). Dijets identified by ESA and JETAN (right).
 } \label{etaphiesajetan}
 \end{center}
\end{figure}

\begin{figure}
\begin{center}
\resizebox{0.7\textwidth}{!}{ \includegraphics{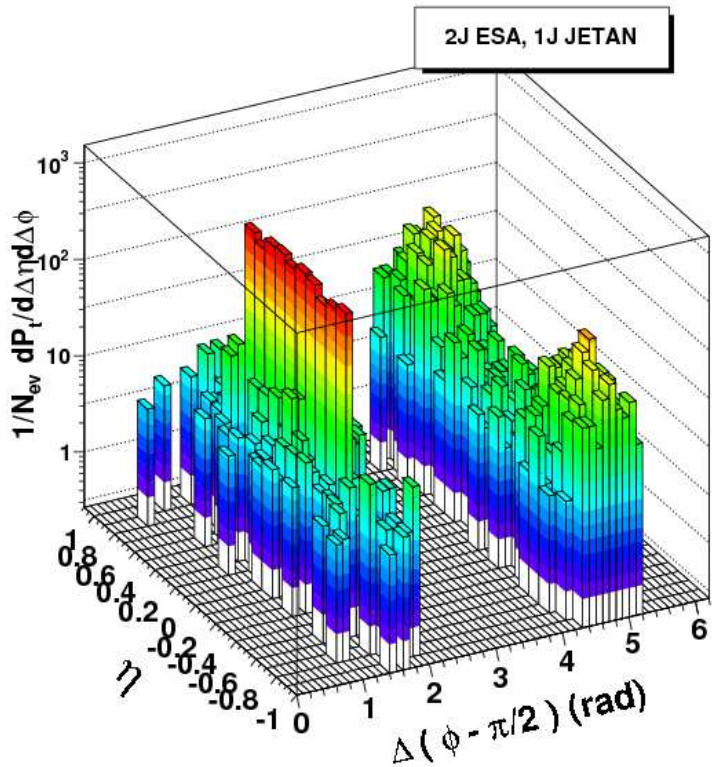} % Please
  }%\vspace{0.5cm}
 \caption{Topologies of the events identified as dijets by ESA but reconstructed as monojet by JETAN. Each bin in $\eta$-$\Delta(\phi - \pi/2)$
 is weighted with the transverse momentum of the charged particles.
 } \label{etaphiesajetan2}
 \end{center}
\end{figure}

The number of jet to be encountered in pp collisions at the LHC
will be very large. In the general case we will find many
overlapping jet cones as is shown in Fig.\ref{overlapping}. On the
other hand the small acceptance for jets causes that many jets are
only partially in the acceptance. We illustrate that with
Fig.\ref{jetan1} where we present the ratio of the difference of
the total transverse momenta found for events  identified by
JETAN\footnote{JETAN is a module of AliRoot which includes
different jet finders. In this analysis we used the UA1 jet finder
algorithm which is based on a cone type\cite{ua1}. The cut in the
transverse momentum of the particles which we included is:
$p_{t}\geq1.5$ in the cone radius: $\sqrt{(\Delta
\phi)^{2}+(\Delta \eta)^{2}}=0.7$. And the minimum transverse
energy of the jets: 1.5 GeV.} as dijets but belonging in the ESA
analysis to two different slices in R. For the lowest one $R<0.2$
we see that the difference of momenta is generally small while for
the slice of the highest R the difference is markedly broader. We
therefore believe that ESA apart from other virtues for the
analysis of bulk properties of the events has a very important
task to play in the identification of specific topologies. The
subsequent comparison of these identified topologies with any kind
of jet finder allows in our mind to study ``well cleaned'' samples
without resorting to different cuts as is usual in the use of the
jet finders. To illustrate our point of view we show two figures:
Fig.\ref{etaphiesajetan}  the distribution of particles in the eta
phi acceptance of ALICE using events identified as dijets by
JETAN, and on the right side the distribution obtained for events
identified as dijets by JETAN applied to ESA dijet events. A
completely clean separation between the near and far side jets is
visible. However, applying JETAN to the ESA dijet sample we found
that a number of events were identified in the present use of ESA
as monojet events. The reason is clear from the eta phi
distribution in Fig.\ref{etaphiesajetan2}  where we plot the
``monojet'' events found by JETAN. The reason is simple - in ESA
we did not limit the eta range for the leading particle of the
``away'' jet. Hence a part of the events are de facto encountered
in the edges of the eta acceptance. The detailed study of these
distribution will be pursued.

\end{document}